\documentclass[printer]{aa}  
\usepackage{graphicx}
\usepackage{txfonts}
\usepackage{rotating}
\usepackage{amssymb}

\newcommand{\kuv}{{\it E\,1821+643}}
\newcommand{\asca}{{\it ASCA}}

\newcommand{\rosat}{{\it ROSAT}}
\newcommand{\xmm}{{\it XMM-Newton}}
\newcommand{\epic}{{\it EPIC}}

\newcommand{\pn}{{\it pn}}
\newcommand{\mos}{{\it MOS}}
\newcommand{\chandra}{{\it Chandra}}

\newcommand{\chicua}{$\chi^2$}

\begin{document}
   \title{XMM-Newton view of the double-peaked Fe K$\alpha$ complex in E1821+643 \thanks{Partially based on observations obtained with XMM-Newton, an ESA science mission with instruments and contributions directly funded by ESA Member States and NASA.}}

\authorrunning{E. Jim\'enez-Bail\'on et al.}
\titlerunning{XMM-Newton view of the double-peaked Fe K$\alpha$ complex in E1821+643}
   \subtitle{}

   \author{E.  Jim\'enez-Bail\'on\inst{1},  M.  Santos-Lle\'o\inst{2},
E.     Piconcelli\inst{3},    G.    Matt\inst{1},     M.    Guainazzi\inst{2},
\and P. Rodr\'{\i}guez-Pascual\inst{2}}

 \offprints{ejimenez@fis.uniroma3.it}

 \institute{Dipartimento di Fisica di l'Universita Roma Tre, Via della Vasca Navle 84, I-00146 Rome, Italy 
\and
XMM-Newton Science Operations Centre, ESAC, ESA, Apartado 50727,
  E-28080 Madrid, Spain
\and
Osservatorio Astronomico di Roma (INAF), Via Frascati 33, I-00040 Monteporzio Catone, Italy
}

\date{Received ; accepted}

 \abstract{We present  the results  of the analysis  of the  hard band
\xmm\           spectra           of           the           luminous,
L$_{2-10keV}\sim3.4\times10^{45}$~erg/s,   radio-quiet  quasar,  \kuv.
Two emission  features were observed  in the 6-7~keV rest  frame band,
confirming  previous  \chandra\  detection  of these  structures.   We
interpret these features as  two single emission lines, one consistent
with  the neutral  Fe K$\alpha$  line at  6.4~keV and  the  other most
likely  due to  FeXXVI.  If  related  to the  quasar, the  high-energy
emission  line should  originate in  highly ionised  matter,  i.e. the
accretion    disc   or    the    clouds   of    the   emission    line
regions. Alternatively, it may  be related to the intergalactic medium
of the  rich galaxy  cluster in which  \kuv\ is embedded.  A composite
broad  emission line  in combination  with an  absorption  line model,
however, also  fits the data  well.  We discuss the  possible physical
interpretations of the origin of these features.

 } \maketitle
%

\section{Introduction}
\label{par:intro}

The  very luminous  radio--quiet quasar  \kuv\ (m$_v$=14.1, z=0.297)
is embedded in  a rich cluster of  galaxies (Hutchings \&  Neff 1991).  A
\rosat\  image of  \kuv\  shows evidence  of  extended X-ray  emission
associated to  the cluster (Hall,  Ellingson \& Green  1997), although
the quasar dominates the X-ray emission.  The core radius of the
cluster      is     $17.\!\!^{\prime\prime}6\pm0.\!\!^{\prime\prime}2$
(1\arcsec=4.4 kpc; Fang et al. 2002).  \kuv\ has been observed extensively in X-rays.  The \asca\ spectrum  confirmed the presence of an emission
line, first discovered  by {\it Ginga} (Kii et  al.  1991), consistent
with  an ionised  state of  iron (Yamashita  et al.   1997). Recently,
\kuv\ was  observed with  both the {\it High  Energy Transmission
Grating Spectrometer (HEGTS) on board \chandra.}  The high-resolution
spectra  show that the  emission line  feature is  complex and  can be
resolved  in a  double-peak  structure.  However,  the origin  of this
feature  is still  not clear.   Fang et  al.  (2002)  argue  that the
feature  is  best  explained  by  two emission  lines,  a  neutral  Fe
K$\alpha$ line, and the other highly ionised, which could originate in a
 highly ionised layer above the  accretion disc.  On the other, Yaqoob
\& Serlemitsos (2005) reanalysed  the data to conclude  that the feature
is a broad iron emission  line with an absorption line superimposed on
the red wing  of the broad line.  The absorption line  could be due to
the resonance  absorption  of highly  ionised  iron,  either in  inflowing
matter  or  in  an  outflow that is strongly  gravitationally  redshifted  and
located very close to the accretion disc.

In this  paper we analyse an  \xmm\ observation of  \kuv. The spectral
analysis (Sect.~3)  was performed in  the 2-10~keV  bandpass and
focused on studying (Sect.~4)  of the double-peak structure that was
also detected in the \xmm\ spectrum.

\section{Observations and data reduction}

The  \xmm\ (Jansen  et al.   2001) observation  of  E1821+643 (Obs-Id.
0110950501) was performed on October   16$^{th}$,  2002.    The   \epic\
observations were performed  in the {\it large window}  (\pn) and {\it
  small  window}  (\mos) using  the  {\it  thin}  filter for  all  
instruments.   The raw  data  were processed  with  the standard  {\it
  Science Analysis  System}, SAS, v.6.5.0 (Gabriel et  al.  2004). The
most updated calibration files available  in October 2005 were used to
process  the data.   According to  the {\it  epaplot}\footnote{The SAS
  task  {\it epatplot}  utilises the  relative ratios  of  single- and
  double-pixel events  that deviate from  standard values in  cases of
  pile-up.} task, no significant  pile-up was detected in any of the \epic\
observations.   We used  WebPIMMS  v3.8a2  to evaluate  the
  expected pile-up  fraction for the measured count  rate and spectral
  shape of \kuv (see next section) and  obtained a pile-up fraction
  smaller  than  0.1 \%.  This   agrees  with  the values  in
the  XMM-Newton     User's     Handbook     (Sect.    3.7.1.3)     \footnote{\tt
    http://xmm.esac.esa.int/external/xmm$\_$user$\_$support/ documentation/uhb}.
The  \epic\  event lists  were  filtered  to  ignore periods  of  high
background flaring according to  the method described in Piconcelli et
al.  (2004).  Due to  the telemetry band, the  \pn\ scientific  buffer was
full several times during the  observation with a result of losing
 time.  The exposures after the filtering are 1.62, 4.52, and 4.57
ks for \pn, {\it MOS-1}, and {\it MOS-2}, respectively.

\section{Spectral analysis} \label{sec:spectral_analysis}

We extracted the  spectra of a circular region  centred on the maximum
emission of  the source and  with 20\arcsec\ and 25\arcsec\  radii for
the  \pn\ and  \mos\ detectors,  respectively. The  background spectra
were extracted from circular  regions of 50\arcsec\ and 15\arcsec 
radius and  located  2\farcm2  and  1\arcmin\  for  \pn\  and  \mos,
respectively, i.e.  outside the galaxy cluster.  In order to apply the
modified \chicua\ minimization technique  (Kendall et al. 1973),
the source \epic\ spectra were  grouped such that each bin contains at
least  20  counts.  The  {\it  MOS-1}  and  {\it MOS-2}  spectra  were
combined to obtain a single spectrum with higher signal--to--noise and
the  \pn\ and \mos\  spectra were  analysed simultaneously.  The count
rates of the spectra in the whole energy band (0.1-10~keV) are 8 and 2
c/s for \pn\ and the {\it MOS1} and {\it MOS2} combined spectra,
respectively.    The  spectral  analysis   was  performed   using  the
XSPEC~v.12.2.0  (Arnaud   1996).   Unless  otherwise   indicated,  all
parameters are given  in the \kuv\ rest frame.   The quoted errors for
the  fit   parameters  refer  to  the  90\%   confidence  level  (i.e.
$\Delta\chi^2=2.71$;  Avni 1976).   We  assumed a  flat $\Lambda  CDM$
cosmology with  ($\Omega_M,\Omega_\Lambda$)=(0.3,0.7) and a  value for
the Hubble constant of 70 kms$^{-1}$ (Bennett et al. 2003).

\subsection{High-energy continuum emission}

The  spectral  analysis was  performed  in  the  2-10\,keV band.   The
Galactic             absorption            (N$_{\rm            H}^{\rm
GAL}=3.8\times10^{20}$\,cm$^{-1}$,   Dickey  \&  Lockman,   1990)  was
included in all the models assuming the photoelectric cross section of
Morrison \& MacCammon  (1983). A simple power law  with a photon index
of  $1.87\pm0.07$  provides  an   acceptable fit  to  the  data  with
\chicua=149 for  177 dof  (degrees of freedom).   The addition  of any
absorption   above   the  Galactic   value   does   not  improve   the
\chicua. However,  residuals are present  in the 4.5-5.5~keV  range in
the observers frame  (5.8-7.1~keV in the quasar rest  frame).  We 
therefore excluded this energy range  in order to better determine the
continuum emission.   We found a power-law photon index  of
$\Gamma=1.92^{+0.09}_{-0.08}$ properly fits  the data (\chicua=112 for
146 dof; model A).  This value of the  photon index is consistent with the mean
value $\sim1.9$  observed for  a large sample  of radio-quiet quasars
(Piconcelli et al. 2005).

\begin{figure}
\includegraphics*[height=90mm,angle=-90]{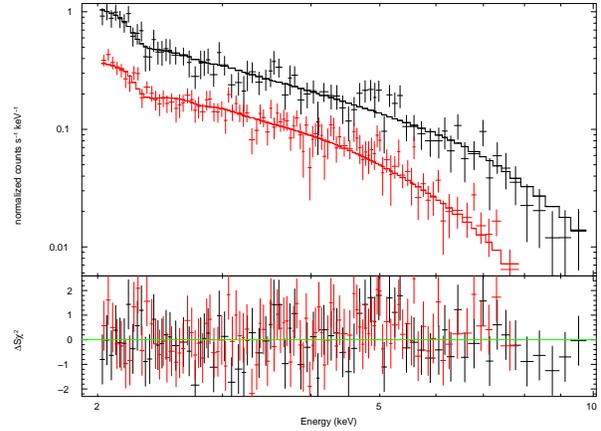}
\caption{ The upper panel  shows the simple power-law  model fitted to
the  \epic\ data, excluding  the 4.5-5.5~keV  band (observed
  energy).
The lower  panel shows the residuals  of the
fitted model to the data  where important residuals are present in the
vicinity of 5~keV  (observed
  energy).}
\label{fig:residuals}
\end{figure}

\begin{figure}
\includegraphics*[width=88mm,angle=0]{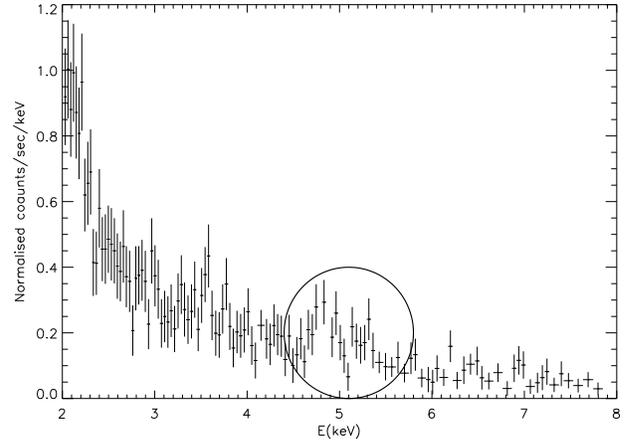}
\caption{The \pn\ spectrum binned to have at least 10 counts
per energy  channel. The plot shows the presence  of a double-peaked
structure located in the vicinity of 5~keV (observed
  energy).}

\label{fig:unbinned}
\end{figure}

Figure~\ref{fig:residuals}  shows   the  residuals  from extrapolating  the
previous fit  in the 4.5-5.5~keV  band, which was  previously ignored.
The deviations  in the  4.5-5.5~keV reveal a  double-peaked structure.
We  also present  (Fig.~\ref{fig:unbinned}) the  \pn\  spectrum binned
with at  least 10  counts-per-channel,  which shows  the double-peaked
structure more  clearly, partially lost  in the 20 counts  per channel
binned spectrum in which just a single point marks the decrease in the
flux  between the two  lines.  In  the 5-6~keV,  i.e. in  the bandpass
where the  residuals are observed, the  effective area of  \pn\ is 2.5
times larger than in the \mos\  detectors. As the \mos\ spectra do not
significantly improve the statistics, we excluded them in the following
fits for simplicity.

\subsection{The iron complex}

 In order to explain  this double feature, we  tested several
models, which are  described in the following.  Table~\ref{tab:models}
shows  these models,  the values  of  the fitting  parameters, and  the
goodness of  each fit.  All these fits were  performed using as the underlying continuum model a   power law with the photon
index  fixed to $\Gamma=1.92$ (model A$_{PN}$).\\

\noindent {\bf Two Gaussian emission lines}

\begin{figure}
\includegraphics*[height=90mm,angle=-90]{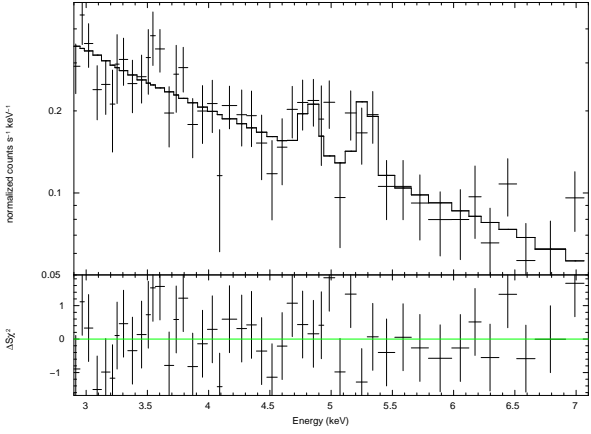}
\caption{ The upper  panel shows  the \pn\ spectrum  and the  best-fit
model, a power law, and two  emission lines. The lower panel shows the
residuals in teerms of sigma of the fitted model to the data.}
\label{fig:bestfit}
\end{figure}

The addition  of a single Gaussian  line fit ($\chi^2=53$  for 76 dof)
results in an  improvement with respect to the  single power-law model
(model A$_{PN}$) with a significance level of $>$99\% according to the
{\it   F-test}.    This   single   Gaussian  line   is   centred   at
$6.28^{+0.4}_{-0.11}$~keV          (rest          frame)          with
$\sigma=130^{+400}_{-90}$~eV.  The inclusion of a second Gaussian line
further  improves  the  fit,  at  a  confidence  level  of  $\sim$98\%
according to  the F-test.  This  double-Gaussian line model  (model B)
provides the simplest model to  account for the residuals .  The lines
are       located        at       $6.27^{+0.18}_{-0.14}$~keV       and
$6.8^{+0.12}_{-0.14}$~keV     with     EW$=120^{+20}_{-50}$~eV     and
EW=$200^{+200}_{-130}$~eV, respectively. The  widths of the lines have
been fixed to  0.01~keV; the fit does not improve  by leaving the
widths  of  the  lines free.   Figure~\ref{fig:bestfit}  shows  this  model,
together with the data and the residuals. \\

\noindent {\bf Relativistic disc line}

\begin{figure}
\includegraphics*[height=90mm,angle=-90]{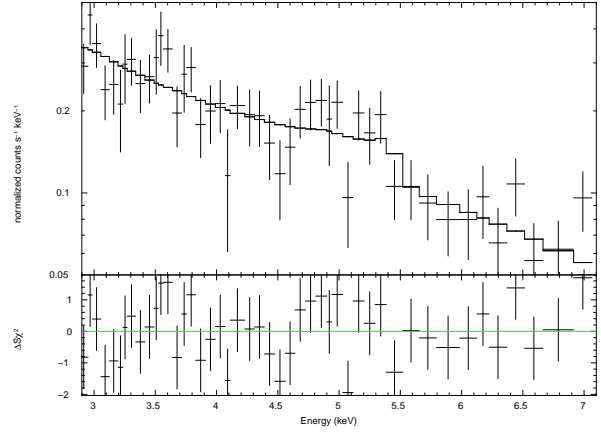}
\caption{ The  upper panel shows  the \pn\ spectrum and  the model
consisting  of a power  law and  a relativistic  disc line.  The lower
panel shows the residuals of the fitted model to the data.}
\label{fig:disc}
\end{figure}

We  also  fitted the  data considering  a relativistic-disc-line
model ({\sl diskline} XSPEC model  Fabian et al. 1989).  We  fixed
the disc inclination to 30$^o$  (Porquet \& Reeves 2003) and the inner
and outer disc  radii to the standard values of  6r$_g$ and 100r$_g$ (
r$_g=GM/c^2$),  respectively. The fit  does not  significantly improve
when the disc inclination and inner and outer disc radii are left free
to  vary.  Moreover,  the  values  of  the  parameters end up   being
unbounded. The fit  (model C, \chicua=55 for 77  dof) improves the fit
with respect to  model A$_{PN}$ with a significance  of 98\% according
to  the F-test.  Even if still acceptable,  the fit is
worse than the one obtained with the two emission Gaussian lines.  The
line energy  is $6.68^{+0.12}_{-0.15}$~keV, which  is incompatible with
the neutral iron ($>$ FeXX).   Moreover, fixing the energy of the line
to 6.4~keV resulted in only  a negligible improvement with respect to a single
power-law  fit.    Figure~\ref{fig:disc}  shows  the  model,  the
observed spectrum, and the residuals.\\

\noindent {\bf Broad emission line plus absorption Gaussian line}

\begin{figure}

\includegraphics*[height=90mm,angle=-90]{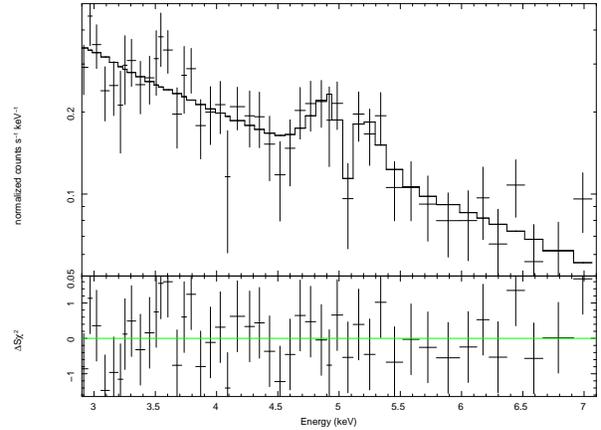}
\caption{The upper  panel shows  the \pn\ spectrum  and  a model consisting of a
 broad emission line and an absorption Gaussian line. The lower panel shows the
residuals of the fitted model to the data.}
\label{fig:absorption}
\end{figure}

According to  Yaqoob \& Serlemitsos  (2005), the preferable  model for
the  HEG spectrum  of \kuv\  includes a  broad emission
line  and a  Gaussian absorption  line  to fit  the residuals  around
5~keV.   We also  applied  this model  (model  D) to  \pn\ data.   The
residuals can be properly fit with this model (\chicua=45 for 73 dof).
The   broad  ($\sigma=240\pm110$~eV)  emission   line  was   found  at
6.53$^{+0.11}_{-0.12}$~keV     and    the    absorption     line    at
6.57$^{+0.05}_{-0.07}$~keV  with  a  $\sigma<180$~eV. Both  lines  are
centred at the same energy, $\sim6.5$~keV, which corresponds to where
the      flux     decreases      between     the      two     emission
features. Figure~\ref{fig:absorption}  shows this model,  together with
the  data and  the residuals.  A similar result  is obtained  when,
instead  of  a  single  Gaussian  line,  a  {\sl  diskline}  model  is
considered for the broad emission line.

\begin{table*}[htb]

\caption{Different models, parameters, and goodness of the fits applied
to the \xmm\ spectrum of \kuv  . {\bf Model A:} Single power-law model
fitted to the  \epic\ data excluding the 4.5-5.5  keV band. Models for
the the  feature around  5~keV applied to  the \pn\ spectrum  with the
photon  index fixed  to  the value  of  model A:  {\bf  Model B:}  Two
emission lines. {\bf Model C:}  Disc line model.  {\bf Model D:} Broad
Gaussian emission line with an absorption line.  {\bf Model E:} Single
Gaussian     line    combined with   a    Raymond-Smith     thermal    plasma
component. The label {\it f} indicates that the parameter has been
fixed to the indicated value.}\label{tab:models}

\begin{tabular}{llllllllllll}
\hline 

 {\bf Model} & $\Gamma$ & \multicolumn{4}{c}{ {\bf Iron Line 1} } & \multicolumn{4}{c}{ {\bf Iron Line 2/Abs. Line} } & {\bf R-S} & {\bf Goodness} \\
 & & E & $\sigma$ & I & EW & E & $\sigma$ & I &EW & kT \\
 & & &  & 10$^{-5}$ & & & & 10$^{-5}$ & \\
& &  keV & keV & ph cm$^{-2}$s$^{-1}$ & eV &  keV & keV & ph cm$^{-2}$s$^{-1}$ & eV &\\
\hline

 A & $1.92^{+0.09}_{-0.08}$ & - & - & - & -  & - & - & - & - & - &112 (146 dof)\\ 
A$_{PN}$ & 1.92 f & - & -  & - & - & - & - &  - & - & - & 61 (79 dof)\\
B & 1.92 f &  $6.27^{+0.18}_{-0.14}$ & $0.01$f & 5$\pm$3 & 120$^{+20}_{-50}$ & $6.85^{+0.12}_{-0.14}$ & $0.01$f & $3\pm2$ & $200^{+200}_{-130}$ & - & 51 (75 dof) \\
C & 1.92 f & $6.68^{+0.12}_{-0.15}$ & - & $6^{+5}_{-4}$ & $400^{+50}_{-200}$ & - & - & - & - & - & 55 (77 dof)\\
D & 1.92 f & $6.53^{+0.11}_{-0.12}$ & $0.24\pm0.11$ & $11^{+26}_{-4}$ &  $700^{+300}_{-400}$  & $6.57^{+0.05}_{-0.07}$ & $<0.18$ & $4^{+30}_{-2}$ & $130^{+900}_{-60}$& - & 45(73 dof)\\
E & 1.92 f & 6.27$\pm0.11$ & $<0.23$ &  $4^{+4}_{-2}$ & 220$^{+90}_{-130}$ & - & - & - & - & $8^{+13}_{-7}$ & 50(74 dof) \\

\hline
\end{tabular}

\end{table*}

\subsubsection{A Possible contamination from the surrounding galaxy cluster?}

\begin{figure}

\includegraphics*[height=90mm,angle=-90]{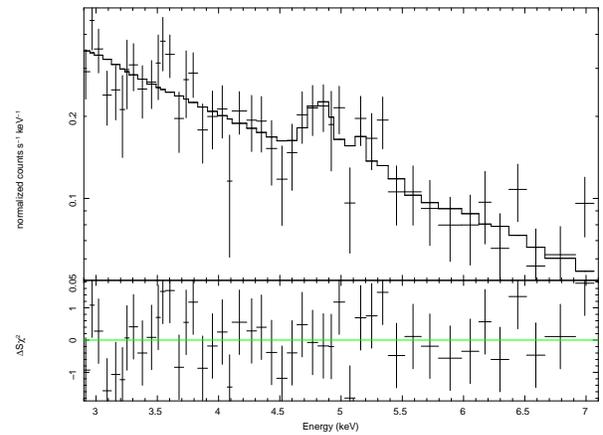}
\caption{The  upper  panel shows  the  \pn\  spectrum  and the model consisting of a  single
emission  line  (E=6.27$\pm$0.11  keV) and the  Raymond-Smith
thermal emission. The  lower panel shows  the residuals  of the
fitted model to the data.}
\label{fig:cluster}
\end{figure}

\begin{figure}
\includegraphics*[height=90mm,angle=-90]{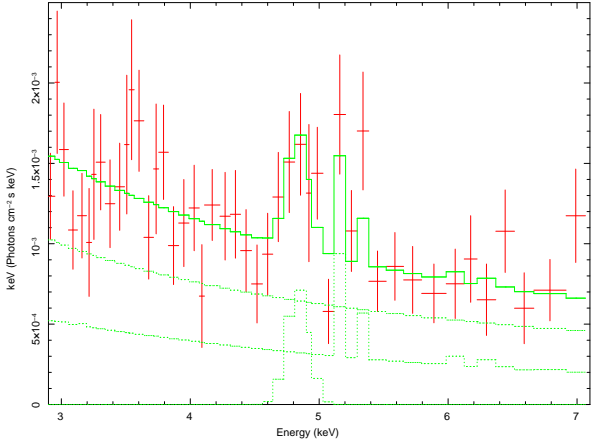}
\caption{The  figure shows  the  unfolded \pn\  spectrum and the models consisting of a 
    single emission line and the Raymond-Smith emission in
units of $Ef(E)$ and the contribution of the various additive components.}
\label{fig:cluster_euf}
\end{figure}

 \kuv\ is known to be embedded in a very rich galaxy cluster. Although
the X-ray images of \kuv\ are clearly dominated by the emission of the
point-like  source,  \rosat\  (Hall,  Ellingson \&  Green  1997),  and
\chandra\  (Fang  et  al.   2002)  images show  evidence  of  extended
emission associated  with the  cluster.  We therefore  investigate the
possibility that the  cluster emission could significantly contaminate
the \kuv\ spectrum.  We then tested  a model consisting of a power law
accounting for  the AGN continuum  emission, a Gaussian  emission line
for the lower energy feature, i.e.  6.27$\pm0.11$~keV, consistent
with neutral  iron combined with a  Raymond-Smith thermal plasma  (Raymond \&
Smith 1977) attributed  to the galaxy cluster emission.   We fixed the
metal  abundance   to  the  typical  value  for   hot  clusters,  i.e.
0.3Z$_\odot$ (Ettori 2005 and references therein).  This value is
also consistent  with the  cluster abundance measuerd  by Fang  et al.
2002  using   \chandra\  data  (Z=0.35$\pm$0.08Z$_\odot$).   The  fit
(\chicua=50 for 74 dof) improves  with respect to the single power law
with  a  single-Gaussian  line  model,  but  the  improvement  is  not
significant according to the F-test, which gives a probability of only
85\%.   Figures~\ref{fig:cluster} and  \ref{fig:cluster_euf}  show the
fitted  model,  the  data  and  the  residuals to  the  fit,  and  the
contribution  of  the  several  components.  The  temperature  of  the
Raymond-Smith  plasma (T=$8^{+13}_{-7}$~keV)  is appropiated  for rich
clusters (Kaastra  et al.  2004), even if  too poorly  constrained for
deriving  any conclusion  in this  respect.  The  line  parameters are
consistent with the values found for the single-Gaussian emission line
and also with the best-fit  parameters of the low-energy emission line
in model B.  The measured  luminosity associated to the thermal plasma
component         in        the        2-10~keV         band        is
$1.1^{+1.2}_{-0.2}\times10^{45}$~erg/s,     which    corresponds    to
30$\pm$6\% of the total luminosity in this band.

\subsection{Flux, luminosity and variability}

Assuming  that the  best-fit  model is  the two  Gaussian  line model
(model  B),  the   \kuv\  flux  observed  in  the   2-10~keV  band  is
$1.24^{+0.04}_{-0.05}\times10^{-11}$~ergs$^{-1}$cm$^{-2}$.         The
corresponding   unabsorbed   luminosity    in   the   same   band   is
$3.44^{+0.13}_{-0.09}\times10^{45}$~erg/s.  Similar  results are found
when the other models are applied.

No sign  of short-term variability  (i.e. during the  observation) has
been  detected.    On  the  contrary,  when   compared  with  previous
observations, a  moderate flux variation has been  measured.  In particular,
the  highest flux  measured in  the 2-10~keV  band corresponds  to the
oldest observation, performed  by \asca\ (June 1993), with  a value of
$(1.7-1.8)\times10^{-11}$~ergs$^{-1}$cm$^{-2}$.      The     \chandra\
observations were performed  3  weeks apart in January and
February      of     2001,  and     the          measured  fluxes     were
$1.1\times10^{-11}$~ergs$^{-1}$cm$^{-2}$                            and
$1.2\times10^{-11}$~ergs$^{-1}$cm$^{-2}$,  respectively.  These values
are     within      the      same      order     of magnitude of     our      result,
$(1.2-1.3)\times10^{-11}$~ergs$^{-1}$cm$^{-2}$,    from    the   \xmm\
observation performed in October 2002.

\section{ Discussion}

Giving the short net \pn\ exposure (only 1.62~ks), resulting from high
background and telemetry bandwidth limitations, the signal to noise of
the \epic\ spectra  does not allow us to discriminate  among the possible
models applied (see  Sect.~\ref{sec:spectral_analysis}) to account for
the double-peaked feature in the 6-7~keV rest frame band.  However, it
is  remarkable  that  even  with  such a  short  exposure  time,  some
interesting results can be still established.

According        to       the        results        described       in
Sect.~\ref{sec:spectral_analysis}, the model consisting of a power law
and    two    narrow    Gaussian    emission    lines    centred    at
$6.27^{+0.18}_{-0.14}$~keV and $6.85^{+0.12}_{-0.14}$~keV provides the
simplest  description  of  the  observed spectrum.  The  two  Gaussian
emission-line  model  provides  acceptable  fits  to  the  spectra  of
previous \chandra\ and \asca\ observations.

The  centroid of  the lower  energy  line ($6.27^{+0.18}_{-0.14}$~keV)
measured in the \pn\ spectrum is compatible within the errors with the
\chandra\  grating  result  ($6.35^{+0.03}_{-0.05}$~keV)  obtained  by
Yaqoob \& Serlemitsos (2005).   This energy is consistent with neutral
iron.  A narrow emission line  centred at $\sim6.4$~keV is observed in
most  \xmm\  spectra of  radio-quiet  quasars (Jim\'enez-Bail\'on  et
al. 2005).  The most likely origin of this feature is material located
in the  outer parts of the  accretion disc, the  molecular {\it torus}
and/or  the Broad  Line  Region. The  value  of the  EW  of this  line
($120^{+20}_{-50}$~eV) also closely matches with the average values found
for  low-luminosity (Yaqoob \&  Padmanabhan 2004)  and high-luminosity
(Jim\'enez-Bail\'on et al. 2005) radio-quiet AGN. It is worth noting
that,      given      the      high     luminosity      of      \kuv\
(L$_{2-10\,keV}\sim3.4\times10^{45}$ erg/s),  this finding is  at odds
with the so-called X-ray Baldwind effect, i.e. the decrement of the EW
as  a function of  luminosity, as  suggested by  Jim\'enez-Bail\'on et
al. (2005), in which no clear correlation between these two quantities
is observed for a large sample of radio-quiet QSO.

The higher  energy line  has an EW  of $200^{+200}_{-130}$~eV  and its
centroid  energy, $6.85^{+0.12}_{-0.14}$~keV,  which is consistent  with the
FeXXV  and FeXXVI lines.  These line  parameters are  fully consistent
with those derived from  previous \chandra\ grating observations (Fang
et  al.  2002  and  Yaqoob  \& Serlemitsos  2005).   This line  should
originate  in  hot  matter.   Therefore, we  considered  two  possible
scenarios for the origin of this  feature: i) the quasar itself or ii)
the contamination from the galaxy  cluster where \kuv\ is embedded. In
the former case,  the feature can be interpreted as  the result of the
illumination of  extended ionised matter, i.e. the  emission line
regions or  the inner walls of  the {\it torus}. It  is also possible
that the ionised  line could originate in the  accretion disc. In this
case,  the  emission  feature  could  correspond to  either  a  narrow
emission line  originated in an ionised  layer on the  disc surface or
the  blue   horn  of  a  relativistically-broadened   iron  line.  The
intensities  of   both  emission  lines  measured   by  \chandra\  are
systematically lower than our \xmm\  values, although due to the large
uncertainties, \xmm\ and  \chandra\ measurements are compatible within
errors.   A {\it  diskline}  model  produced a  statistically
acceptable  fit (\chicua=55  for  77 dof)  to  the double-peaked  iron
complex in  the \pn\ spectrum of  \kuv.  The line energy  was found at
$6.68^{+0.12}_{-0.15}$~keV, so  compatible  with  an origin  from  ionised
matter ($\geq$ FeXXI).

We  also  investigated the  possibility  that  the surrounding  galaxy
cluster could significantly contaminate  the source spectrum. The \pn\
extraction radius  of 20$^{\prime\prime}$ includes the  entire core of
the  cluster,  i.e. $~17.\!\!^{\prime\prime}6$  (Fang  et al.   2002).
Therefore we added a Raymond-Smith thermal plasma component attributed
to the diffuse cluster emission to the power-law with an emission-line
($6.27\pm0.11$~keV)  model .   The metal  abundance was  fixed  to the
typical  value of  hot clusters,  i.e.  0.3Z$_\odot$.   The parameters
(temperature  and  X-ray  luminosity)  derived for  the  Raymond-Smith
component  are  consistent with  the  typical  values  of rich  galaxy
clusters.  In  particular, the  temperature ($kT\sim8$~keV) is  in the
range found for hot galaxy clusters.  The luminosity associated to the
thermal  component,  L$_{\rm X}=1.1^{+1.2}_{-0.2}\times10^{45}$~erg/s,
is consistent with the temperature-luminosity relationship for cluster
cores, L$_{\rm X}\sim5\times10^{42}(kT)^{3}$~erg/s (Ota et al.  2006).
Fang et  al. (2002) have found  that the radial profie  of the diffuse
emission associated to the cluster can be modeled by a $\beta$-profile
(Cavaliere \&  Fusco-Femiano 1976). Adopting their  parameters, we find
that in  the 20''  \xmm\ extraction region,  around 50\% of  the total
cluster emission  would be included, corresponding to  a luminosity of
$1.4\times10^{45}$  erg/s, which is fully consistent  with our  measurement. On
the other  hand, given the  measured temperature of  the Raymond-Smith
component,  a  significant  contribution  from the  FeXXV  and  FeXXVI
K$\alpha$ lines is expected.  In fact, the measured flux for these two
emission                            lines                           is
$(1.8\pm1.7)\times10^{-5}$~photons\,cm$^{-2}$\,s$^{-1}$.  By comparing
this value  with the line flux of  the $6.85^{+0.12}_{-0.14}$~keV line
inferred                  by                  model                  B
(i.e. $(2.6^{+1.6}_{-0.5})\times10^{-5}$~photons\,cm$^{-2}$\,s$^{-1}$),
the  hot plasma  emission can  in  principle account  for the  6.8~keV
emission  feature.   However,  the  large  errors  indicate  that  the
contribution  of the  cluster to  the high-energy  emission  line flux
could be as  low as of $\sim$2\%.  Interestingly,  Fang et al.  (2002)
estimated the contribution  of the cluster to be  between 3\% and 50\%
of  the flux  line,  when only  the  HEG spectrum  is considered,  and
between 30\% and 100\% for the combined, higher signal--to--noise, HEG
and MEG spectra.\\

Finally, another  possible origin for the  double-peaked iron emission
is a combination of a  broad emission line and an absorption line.  This
scenario was suggested  by Yaqoob  \& Serlemitsos  (2005) who
argue that the absorption line could be attributed to redshifted resonance absorption by  highly  ionised  material,   most  likely  an  inflow.   The  high
ionisation state of the iron is supported by the absence of soft X-ray
absorption features  (Yaqoob \& Serlemitsos 2005; Mathur  et al. 2003;
Fang et al.  2002). Matt  (1994) suggests for the first time the
importance of the effect of resonant K$\alpha$ transitions due to iron
ions that can absorb continuum photons in the $\sim6.4-6.7$~keV band.
Ruszkowski \&  Fabian (2000) computed the absorbing  effect of ionised
iron  located  in  a  rotating  diffuse  plasma  that surrounds  the
accretion  disc and the  corona.  The  FeXXV and  FeXXVI can
resonantly  absorb photons  of the  continuum and  of  the fluorescent
K$\alpha$  emission line  emitted from  the primary  source,  i.e. the
corona and the  accretion disc.   Some  observational  evidence of
absorption lines due  to ionised iron has been  reported: e.g. Dadina
et al.   (2005)  observed in  \xmm\ and {\it  BeppoSAX} spectra of
Mrk~509 red-  and blue-shifted absorption  lines due to  ionised iron;
the  \xmm\  spectrum  of  Q0056-363  shows  evidence  of  a  transient
absorption  line  interpreted  as  redshifted ionised  iron  (Matt  et
al. 2005);  and recently, Markowitz  et al. 2006  detected a  blue-shifted
absorption FeXXVI  K$\alpha$ line in IC~4329A, suggesting the presence  of an ionised
outflow component. In  our analysis,  the
absorption  line was  found at  $6.57^{+0.05}_{-0.07}$~keV, consistent
with Fe  states from XX to  XXII.  This value  is largely incompatible
with   the   one   measured   by   Yaqoob   \&   Serlemitsos   (2005),
$6.228^{+0.018}_{-0.013}$~keV. Fixing the  values of the parameters of
both emission and  absorption lines in our model  to the ones obtained
by  Yaqoob  \& Serlemitsos  (2005)  leads  to  an unsatisfactory  fit.
However,  Yaqoob   \&  Serlemitsos  (2005)   indicate  that  the
absorption line could be variable and therefore would not be observed
at   the  same   energy  (or   even   unpresent  at   all)  in   other
observations. If this  is the case, and assuming that  the line is due
to  the  K$\alpha$  transition  from  FeXXVI, then  it  would  be
redshifted with a  velocity of 17200$^{+3000}_{-2600}$~km/s, $\sim$6\%
of the speed  of light.

\section{Conclusions}

In this  paper, we have presented the  analysis of the high-energy band of
the      \pn\      \xmm\      spectrum      of      the      luminous,
L$_{2-10keV}=3.44^{+0.13}_{-0.09}\times10^{45}$~erg/s,   radio-quiet
quasar  \kuv.  The  spectrum  shows the  presence  of a  double-peaked
feature  observed  in  the  6-7~keV  quasar  rest  frame  band.   This
structure was  also detected  in two \chandra\  observations performed
with the HEG  and LEG.  Our analysis of  the \xmm\ data therefore
confirms the  results of previous  observations and also  provides new
possible  physical interpretations  of  the origin  of these  emission
features.

Several models have been
tested to fit the  emission properties.   The  simplest model  corresponds  to  a  couple of  Gaussian
emission  lines.  According  to  our interpretations,   the  lower
energy  line  is  consistent  with  the narrow  neutral  Fe  K$\alpha$
line.  Within the errors,  the calculated  parameters of  the emission
lines  are  compatible with  previous  \chandra\ observations.  Matter
located in the outer parts  of the accretion disc, the molecular torus,
and/or the the Broad Line Region is likely to be responsible of the emission
line.

On  the  other  hand,  the  centroid  of the  higher  energy  line  is
compatible  with the K$\alpha$  transition from  FeXXV and  FeXXVI and
therefore it had to originate  in matter that was hot enough. Assuming
that  the line originated  in the  quasar itself,  the feature  can be
interpreted as a narrow emission  line originated in a highly ionised
atmosphere  of the  accretion  disc,  as the  blue-shifted  peak of  a
relativistic emission  line or as  narrow emission line caused  by the
illumination of  extended photoionised matter, i.e.  the emission
line regions or the inner  parts of the {\it torus}.  The statistical
analysis  is slightly  unfavourable to   the relativistically  broaden  emission
line.  We have also considered  the possibility that the higher energy
line is due  to contamination by the rich cluster  in which the source
is  embedded.  The  measured  temperature of  the  hot diffuse  plasma
attributed to  the cluster,  kT=$8^{+13}_{-7}$~keV, is high  enough to
produce an  important FeXXVI emission  line.  The \xmm\ data  analysis has
revealed  for  the  first  time  that the  predicted  plasma  emission
attributed to the cluster could fully explain the measured flux of the
high-energy emission feature.

The  line complex  may  be also  be  interpreted as  a broad  Gaussian
emission line, compatible with  the neutral iron combined with an absorption line,
observed at $6.57^{+0.05}_{-0.07}$~keV. This model was first suggested
by Yaqoob \& Serlemitsos (2005)  to explain the \chandra\ HEG
data.  If this model is correct, the centroid energy of the absorption
line must be variable, changing from $\sim$6.23~keV in the HEG data to
$\sim$6.57~keV in  our \xmm\ spectrum, thus indicating  a large variability
of the  physical properties  (location, velocity...) of  the absorbing
matter.

Neither short-term variability, i.e.   within the \xmm\ observation, nor
mid-term,   i.e.  less   than   22  months,    has   been
detected.   Although   a   decrease   in   flux  of   the   order   of
$\sim5\times10^{-12}$~ergs$^{-1}$cm$^{-2}$ has been detected when comparing
our observed  flux (October 2002)  with that measured by  \asca\ (June
1993).

Unfortunately,  given the  short net  exposure time  results on  a low
signal--to--noise spectrum, the \xmm\ data prevented our making  firm
conclusions on  the real nature of  the double-peaked feature
in \kuv. Long enough  \xmm\ observations of the source  would be 
valuable  for determining both the  best among several   tested models  and
the values  of the parameters,  in order  to shed light on the nature of the X-ray \kuv\ emission.

\acknowledgements{The  authors  would  like  to  thank  the  anonymous
referee for the very  useful comments, that significantly improved the
paper. We also  thank Nobert Schartel for very  useful discussions. EJ
and    MSL    acknowledge    funding    from   Spanish    MEC    grant
AYA2004-08260-C03-03.}\\

\end{document}